\documentclass[12pt,showpacs,amsmath,amssymb]{revtex4}

\usepackage{graphicx}
\usepackage{dcolumn}
\usepackage{bm}

\newcommand{\AU}{\operatorname{AU}}

\begin{document}

\title{Nonlocal Electrodynamics of Rotating Systems}

\author{Bahram Mashhoon}
\affiliation{Department of Physics and Astronomy\\
University of Missouri-Columbia\\
Columbia, Missouri 65211}

\begin{abstract} The nonlocal electrodynamics of uniformly rotating systems is presented and its predictions are
discussed. In this case, due to paucity of adequate experimental data, the nonlocal theory cannot be directly 
confronted with observation at present. The approach adopted here is therefore based on the correspondence principle: 
the nonrelativistic quantum physics of electrons in circular ``orbits" is studied. The helicity dependence of 
the photoeffect from the circular states of atomic hydrogen is explored as well as the resonant absorption of 
 a photon by an electron in a circular ``orbit" about a uniform magnetic field. Qualitative agreement of the 
predictions of the classical nonlocal electrodynamics with quantum-mechanical results is demonstrated in the 
correspondence regime.\end{abstract}

\pacs{03.30.+p, 11.10.Lm, 03.65.Sq, 04.20.Cv}

\maketitle

\section{\label{s1}Introduction}

Imagine a background inertial reference frame with coordinates $x^\mu =(t,{\bf x})$ in Minkowski spacetime. An
accelerated observer follows a worldline $x^\mu (\tau )$, where $\tau$ is the proper time along its path. According to
the standard theory of relativity, the accelerated observer---at each instant along its worldline---is equivalent to
an otherwise identical momentarily comoving inertial observer \cite{1,2}. This {\it hypothesis of locality}
implies that an accelerated observer passes through a continuous infinity of hypothetical momentarily comoving inertial
observers. Lorentz invariance can then be locally extended to non-inertial observers via this basic assumption. Each
inertial observer is endowed with an orthonormal tetrad frame; therefore, the hypothesis of locality implies that an
accelerated observer carries an orthonormal tetrad frame $\lambda ^\mu _{\;\;(\alpha )}(\tau )$ along its worldline
such that $\lambda^\mu_{\;\; (0)}=dx^\mu /d\tau $ is its temporal axis and $\lambda^\mu _{\;\;(i)}$, $i=1,2,3,$ are
the unit axes of its local spatial frame.

The motion of the tetrad frame along the worldline of the accelerated observer may be expressed as
\begin{equation}\label{eq1} \frac{d\lambda^\mu_{\;\;(\alpha)}}{d\tau
}=\Phi_\alpha^{\;\;\beta}\lambda^\mu_{\;\;(\beta)},\end{equation}
where $\Phi_{\alpha\beta}(\tau)$ is the antisymmetric acceleration tensor. In analogy with the Faraday tensor,
$\Phi_{\alpha\beta}\to (-\tilde{\bf g}, \tilde{\boldsymbol{\Omega}})$; that is, the ``electric" part
$(\Phi_{0i}=\tilde{g}_i)$ consists of the translational acceleration of the observer, while its ``magnetic" part
$(\Phi_{ij}=\epsilon _{ijk}\tilde{\Omega}^k)$ consists of the rotational frequency of the spatial frame with respect to
a nonrotating (i.e. Fermi-Walker transported) tetrad frame. The scalar invariants $\tilde{\bf g}(\tau )$ and
$\tilde{\boldsymbol{\Omega}}(\tau )$ characterize the rate of change of the state of the observer and may be used to
construct the acceleration scales---i.e. acceleration length $\mathcal{L}$ and acceleration time $\mathcal{L}/c$---of
the observer. For an Earth-based laboratory, for instance, $\mathcal{L}=c^2/g_\oplus\cong
$ 1 lt-yr is the translational acceleration length, while $\mathcal{L}=c/\Omega_\oplus
\cong 28 \AU$ is the rotational
acceleration length.

If all physical phenomena could be reduced to pointlike {\it coincidences}, then the hypothesis of locality would be
strictly valid. Indeed, the hypothesis of locality originates from Newtonian mechanics, where the state of a point
particle is characterized by its position and velocity. The accelerated particle and the momentarily comoving inertial
particle have the same state; therefore, they are pointwise physically equivalent. Thus no new physical hypothesis is
needed in the Newtonian treatment of accelerated systems. However, for
wave phenomena we expect  deviations from the hypothesis of locality that would be
proportional to $\lambda /\mathcal{L}$, where $\lambda$ is the wavelength of the
radiation. To illustrate this viewpoint, consider the measurement of the frequency of
incident radiation by an accelerated observer. At least a few oscillations of the incident
wave must be received by the observer before a reasonable determination of its frequency
can be made; however, during this time interval of $\sim \lambda /c$ the state of the
observer has changed. Nevertheless, this may be ignored if $\lambda$ is sufficiently
small compared to $\mathcal{L}$. The consistency of this approach can be illustrated in
the case of a classical particle of mass $m$ and charge $q$ that is accelerated by an
external force  ${\bf f}$. The accelerated charge radiates electromagnetic radiation of
wavelength $\lambda \sim
\mathcal{L}$,  where $\mathcal{L}$ is the particle's acceleration length. It is expected
that the hypothesis of locality is violated in the interaction of the particle with the
electromagnetic field as $\lambda /\mathcal{L}\sim 1$. Thus the state of the particle
cannot be characterized by its position and velocity as demanded by the hypothesis of
locality. This is in agreement with the equation of motion of the particle, which in the
nonrelativistic approximation may be expressed as
\begin{equation}
m\frac{d{\bf v}}{dt}-\frac{2}{3} \frac{q^2}{c^3} \frac{d^2{\bf v}}{dt^2}+\dots ={\bf
f}.\label{eq2}\end{equation}
The dependence  of this Abraham-Lorentz equation on the temporal derivative of the
acceleration implies that the state of a radiating particle cannot be adequately
characterized by its position and velocity.
 
It follows from these arguments that it is necessary to
contemplate a generalization of the hypothesis of
locality that would be adequate for all wave phenomena. To proceed, we restrict our
attention to the measurement of an electromagnetic radiation field $F_{\mu\nu}$ by an
accelerated observer, though the general approach is applicable to any radiation field~\cite{3}.
According to the hypothesis of locality, the accelerated observer
may be replaced at each instant by the momentarily comoving inertial observer for which
the measured field is equivalent to the projection of $F_{\mu\nu}$ onto its tetrad
frame. Consider the class of fields $F_{(\alpha )(\beta)}(\tau )$ measured pointwise by
the hypothetical momentarily comoving inertial observers; then,
\begin{equation}\label{eq3} F_{(\alpha )(\beta)}(\tau )=F_{\mu\nu}\lambda^\mu
_{\;\;(\alpha)} \lambda^\nu _{\;\;(\beta)}.\end{equation}
Let $\mathcal{F}_{(\alpha )(\beta)} (\tau )$ be the electromagnetic
field that is
actually
measured by the accelerated observer. The hypothesis of locality states that at each
instant $\tau$, $\mathcal{F}_{(\alpha )(\beta)}(\tau )$ and $F_{(\alpha )(\beta)}(\tau
)$ are the same. On the other hand, the most general linear relationship between
$\mathcal{F}_{(\alpha )(\beta)}(\tau )$ and $F_{(\alpha )(\beta)}(\tau )$ consistent
with causality is \cite{3}
\begin{equation}\label{eq4} \mathcal{F}_{(\alpha )(\beta)}(\tau )=F_{(\alpha )(\beta )}
(\tau )+u(\tau -\tau_0)\int^\tau_{\tau_0}K_{(\alpha
)(\beta)}^{\;\;\;\;\;\;\;\;\;\;(\gamma
)(\delta)} (\tau ,\tau ')F_{(\gamma )(\delta)} (\tau ')d\tau ',\end{equation}
where $u(x)$ is the unit step function such that $u(x)=1$ for $x>0$ and $u(x)=0$ for
$x<0$. The kernel $K$ is expected to be directly related to the acceleration of the
observer. In Eq.~\eqref{eq4}, the measured field involves a weighted average over the past
worldline of the observer. This is consistent with the viewpoint developed by Bohr and
Rosenfeld \cite{4} that a pointwise field determination is not possible in principle and
must indeed be replaced by a certain averaging process. Using the decomposition
$F_{\mu\nu}\to ({\bf E},{\bf B})$, it is useful to replace $F_{\mu\nu}$ by a column
$6$-vector $F$ that has ${\bf E}$ and ${\bf B}$ as its components, respectively. Thus
Eq.~\eqref{eq3} may be expressed as $\hat{F}=\Lambda F$, where $\Lambda (\tau )$ is a
$6\times 6$
matrix and is a representation of the Lorentz group. In this way, Eq.~\eqref{eq4} may be
written as
\begin{equation}
\label{eq5} \hat{\mathcal{F}}(\tau )=\hat{F}(\tau )+u(\tau -\tau_0)\int^\tau_{\tau_0}
\hat{K} (\tau ,\tau')\hat{F}(\tau ')d\tau '.
\end{equation}
The Volterra integral equation \eqref{eq5} implies, via Volterra's theorem~\cite{5},
that in the space of continuous functions the relationship between $\hat{\mathcal{F}}$
and $F$ is unique. Volterra's theorem has been extended to the Hilbert space of
square-integrable functions by Tricomi~\cite{6}. The ansatz \eqref{eq4} is manifestly
Lorentz covariant; moreover, the kernel is given by quantities that are invariant under
Poincar\'e transformations of the background Minkowski spacetime.

To determine the kernel $\hat{K}$, we postulate that an electromagnetic radiation field
can never stand completely still with respect to an accelerated observer~\cite{3}. This is
a simple generalization of a well-known result of Lorentz invariance, which may be
illustrated using the Doppler effect. If $\omega$ and ${\bf k}$ are the frequency and
wave vector of an electromagnetic wave with respect to static inertial observers, then
an inertial observer moving with velocity ${\bf v}$ would measure a frequency $\omega
'=\gamma (\omega -{\bf v}\cdot {\bf k})$. This implies that $\omega '=0$ if and only if
$\omega =0$. Imposing the same requirement in the general noninertial case, we conclude
that if $\hat{\mathcal{F}}$ turns out to be constant in time, then $F$ must have been a
constant field in the first place. This requirement leads to an integral equation that
could be solved in principle to determine $\hat{K}$~\cite{3}. The Volterra-Tricomi
uniqueness theorem then implies that for any realistic radiation field $F$, the measured
field $\hat{\mathcal{F}}$ will definitely depend upon time. A detailed analysis
\cite{7,8,9} reveals that the unique kernel of the nonlocal theory of accelerated
observers is given by $\hat{K}(\tau ,\tau ')=\hat{k}(\tau ')$, where
\begin{equation}\label{eq6} \hat{k}(\tau ')=-\frac{d\Lambda (\tau ')}{d\tau '}
\Lambda^{-1} (\tau ').\end{equation}

Some of the observational consequences of this nonlocal theory of accelerated observers
have been worked out in the case of linearly accelerated observers \cite{10};
therefore, the present paper is devoted to the observational consequences of the theory
for rotating observers. These are worked out in section~\ref{s2} for a uniformly
rotating observer. Indirect and qualitative evidence in support of the theory is
presented in sections~\ref{s3}, \ref{s4} and \ref{s5}. Section~\ref{s6} contains a
discussion of our results. Mathematical details are relegated to the appendices.

\section{\label{s2}Rotation-Induced Nonlocality}

We now consider the application of the nonlocal theory of accelerated systems to the
important special case of a uniformly rotating observer in the $(x,y)$ plane. Let us
assume that for $-\infty <t<0$, the observer moves along the straight line parallel to
the $y$ axis at $x=r>0$ with constant speed $v$ and arrives at $x=r$ and $y=0$ at $t=0$.
From this instant on, it is forced to move in the positive sense on a circle of radius
$r$ with uniform frequency $\Omega =v/r$. The  azimuthal angle that indicates the
position of the observer for $t\geq 0$ is given by $\varphi =\Omega t=\gamma \Omega
\tau$, where $\gamma$ is the Lorentz factor corresponding to $\beta=v/c=r\Omega /c$. For
$t\geq 0$, the natural tetrad frame of the observer is given by
\begin{eqnarray}\label{eq7} \lambda^\mu_{\;\;(0)} & =& \gamma (1,-\beta \sin \varphi
,\beta \cos \varphi ,0),\nonumber\\
\lambda^\mu_{\;\;(1)}&=& (0,\cos \varphi ,\sin \varphi , 0),\nonumber\\
\lambda^\mu_{\;\;(2)}&=&\gamma (\beta , -\sin \varphi , \cos \varphi ,0),\nonumber\\
\lambda^\mu_{\;\;(3)}&=& (0,0,0,1),\end{eqnarray}
with respect to the global inertial coordinates $(t,x,y,z)$. In this case,
Eq.~\eqref{eq3} implies that
\begin{equation}\label{eq8} \Lambda = \begin{bmatrix} \Lambda_1 & \Lambda_2\\ -\Lambda_2
& \Lambda_1\end{bmatrix},\end{equation}
where
\begin{equation}\label{eq9} \Lambda_1=\begin{bmatrix} \gamma \cos \varphi & \gamma \sin
\varphi & 0\\ -\sin \varphi & \cos \varphi & 0\\ 0 & 0 & \gamma \end{bmatrix},\quad
\Lambda_2=\beta \gamma \begin{bmatrix} 0 & 0 & 1\\ 0 & 0 & 0\\-\cos \varphi & -\sin
\varphi & 0\end{bmatrix}.\end{equation}
Moreover, it follows from Eq.~\eqref{eq1} that the observer has invariant translational
centripetal acceleration
$\tilde{\bf g}=-v\gamma^2\Omega  (1,0,0)$ and rotational frequency
$\tilde{\boldsymbol{\Omega}}=\gamma^2\Omega (0,0,1)$ that are defined with respect to
the spatial axes $\lambda^\mu_{\;\; (i)}$, $i=1,2,3$, that correspond to the radial, tangential and $z$ directions,
respectively. In terms of
these components of the
acceleration tensor, the kernel \eqref{eq6} turns out to be a constant $6\times 6$
matrix given by
\begin{equation}\label{eq10} \hat{k}=\begin{bmatrix}\tilde{\boldsymbol{\Omega}}\cdot
{\bf I} & -\tilde{\bf g}\cdot {\bf I}\\
\tilde{\bf g}\cdot {\bf I} & \tilde{\boldsymbol{\Omega}}\cdot {\bf
I}\end{bmatrix},\end{equation}
where $I_i$, $(I_i)_{jk}=-\epsilon_{ijk}$, is a $3\times 3$ matrix proportional to the
operator of infinitesimal rotations about the $x^i$ axis. It follows that the
electromagnetic field as measured by the rotating observer for $t\geq 0$ is given by
\begin{eqnarray}\label{eq11} \mathcal{E}_1&=& \gamma (\cos \varphi E_1+\sin \varphi
E_2)+\beta \gamma B_3+\gamma^2\Omega \int^\tau_0 (\sin \varphi 'E_1-\cos \varphi
'E_2)d\tau',\nonumber\\
\mathcal{E}_2&=& -\sin \varphi E_1+\cos \varphi E_2+\gamma \Omega \int^\tau_0(\cos
\varphi 'E_1+\sin \varphi 'E_2)d\tau ',\nonumber\\
\mathcal{E}_3&=& \gamma E_3-\beta \gamma (\cos \varphi B_1+\sin \varphi B_2)+\beta
\gamma^2\Omega \int^\tau_0(-\sin \varphi 'B_1+\cos \varphi 'B_2)d\tau ',\nonumber\\
\mathcal{B}_1 & =& \gamma (\cos \varphi B_1+\sin \varphi B_2)-\beta \gamma
E_2+\gamma^2\Omega \int^\tau_0(\sin \varphi 'B_1-\cos \varphi 'B_2)d\tau ',\nonumber\\
\mathcal{B}_2 &=& -\sin \varphi B_1 +\cos \varphi B_2+\gamma \Omega \int^\tau_0(\cos
\varphi 'B_1+\sin \varphi 'B_2)d\tau ',\nonumber\\
\mathcal{B}_3 & = & \gamma B_3+\beta \gamma (\cos \varphi E_1+\sin \varphi E_2)+\beta
\gamma^2\Omega \int^\tau _0 (\sin \varphi 'E_1-\cos \varphi 'E_2)d\tau '.\end{eqnarray}

Let us next consider a normally incident plane monochromatic wave of frequency $\omega$
given by
\begin{equation}\label{eq12} F_\pm (t,{\bf x})=i\omega a\begin{bmatrix} {\bf e}_\pm \\
{\bf b}_\pm \end{bmatrix} e^{-i\omega (t-z/c)},\end{equation}
where $a$ is a complex amplitude ${\bf e}_\pm =\mp (\hat{\bf x}\pm i\hat{\bf y})/\sqrt{2}$,
${\bf b}_\pm =\mp i{\bf e}_\pm$ and the upper (lower) sign represents positive
(negative) helicity radiation. Here, ${\bf e}_\pm$ and ${\bf b}_\pm$ are unit circular
polarization vectors such that ${\bf e}^\ast_\pm =\mp {\bf e}_\mp$ and ${\bf e}_\pm
\cdot {\bf e}^\ast_\pm =1$. As usual, we define the intensity of the radiation field
\eqref{eq12} to be $I_0=\frac{1}{2}\omega^2|a|^2$. In employing complex fields, such as
in Eq.~\eqref{eq12}, we adopt the convention that only the real part of the field is of
physical interest. This is compatible with our general linear approach based on the
superposition principle.

Along the worldline of the rotating observer $z=0$ and $t=\gamma \tau$ in
Eq.~\eqref{eq12}, so that from $\hat{F}=\Lambda F$ we obtain in accordance with the
hypothesis of locality 
\begin{equation}\label{eq13} \hat{F}_\pm (\tau )=i\gamma \omega a\begin{bmatrix}
\hat{\bf e}_\pm \\ \hat{\bf b}_\pm \end{bmatrix} e^{-i\hat{\omega} \tau},\end{equation}
where $\hat{\bf b}_\pm  =\mp i\hat{\bf e}_\pm$ and
\begin{equation}\label{eq14} \hat{\bf e}_\pm =\mp \frac{1}{\sqrt{2}}\begin{bmatrix} 1 \\
\pm i\gamma^{-1}\\ \pm i\beta \end{bmatrix}\end{equation}
are unit vectors such that $\hat{\bf e}_\pm \cdot \hat{\bf e}^\ast_\pm =1$ and $\hat{\bf
e}_\pm \to {\bf e}_\pm$ as $\beta \to 0$. Moreover, $\hat{\omega} =\gamma (\omega \mp
\Omega)$ in Eq.~\eqref{eq13}, so that the transverse Doppler effect is modified by the
helicity-rotation coupling. The observational evidence in support of the
helicity-rotation coupling for $\omega \gg \Omega$ is discussed in \cite{11,12,13}. The
general spin-rotation-gravity coupling has been reviewed in \cite{14}.

It follows from Eq.~\eqref{eq11} that the nonlocal field measured by the rotating
observer is given by (cf. Appendix~\ref{aA})
\begin{equation}\label{eq15} \hat{\mathcal{F}}_\pm =i\gamma \omega a\begin{bmatrix}
\hat{\bf e}_\pm \\ \hat{\bf b}_\pm \end{bmatrix} \frac{e^{-i\hat{\omega}\tau}\mp
\frac{\Omega}{\omega}}{1\mp \frac{\Omega}{\omega}}.\end{equation}
It is important to recognize two novel aspects of this nonlocal result as compared to
the local result given by Eq.~\eqref{eq13}: (i) The oscillatory part of Eq.~\eqref{eq15}
is the same as in Eq.~\eqref{eq13} except for the multiplicative factor of $(1\mp \Omega
/\omega )^{-1}$. It follows from this feature that the measured intensity of the
positive helicity incident wave with $\omega >\Omega$ is enhanced by a factor of
$(1-\Omega /\omega)^{-2}$, while the corresponding negative helicity intensity is
diminished by a factor of $(1+\Omega /\omega )^{-2}$. Thus for the same incident
frequency $\omega$ and intensity $I_0$, the ratio of the measured intensity of the
positive helicity radiation $I_+$ to the measured intensity of the negative helicity
radiation $I_-$ is
\begin{equation}
\label{eq16} \frac{I_+}{I_-}=\left( \frac{\omega +\Omega}{\omega
-\Omega}\right)^2.\end{equation}
(ii) In contrast with Eqs.~\eqref{eq12} and \eqref{eq13}, Eq.~\eqref{eq15} contains a
constant part proportional to $\Omega$; in fact, this constant term is necessary in
order that $\hat{\mathcal{F}}_+(\tau )$ would have a proper limit in the resonance case
$\omega =\Omega$. Indeed, for $\omega =\Omega$ we have $\hat{\omega}=0$ in the positive
helicity case, $\hat{\omega}=2\gamma \Omega$ in the negative helicity case and
\begin{equation}\label{eq17} \hat{\mathcal{F}}_\pm (\tau )=i\gamma \Omega
a\begin{bmatrix} \hat{\bf e}_\pm \\ \hat{\bf b}_\pm \end{bmatrix} f_\pm (\tau
),\end{equation}
where $f_+=1-i\gamma \Omega \tau$ and $f_-=\cos (\gamma \Omega \tau)\exp (-i\gamma
\Omega \tau )$. Thus in the positive helicity case the measured field grows indefinitely
with proper time. This linear divergence of the field is a consequence of the fact that
the incident plane wave has a constant amplitude for all time; that is, the divergence
would disappear for any finite incident wave packet. Let us note that in the
corresponding negative helicity case, the measured intensity is given by
$I_-=\frac{1}{2}\gamma^2I_0$. It follows that for $\omega=\Omega$, $I_+/I_-\to \infty$
as $\Omega \tau \to \infty$.

These predictions of the nonlocal theory of accelerated observers follow {\it directly}
from the {\it nonlocality} of our ansatz~\eqref{eq4}. It is therefore important to provide
observational evidence for these predictions.

Available data regarding rotating systems involve situations with $\omega \gg \Omega$
and appear to be consistent with $I_+\approx I_-$ and the nonlocal theory within the
limits of the accuracy of the observations~\cite{13}. Imagine, for instance, radio waves
of frequency 10 GHz normally incident on a disk that is rotating very rapidly at a rate
of $10^3$ rounds per second; then, $\Omega /\omega =10^{-7}$ and $I_+/I_-\approx
1+4\Omega /\omega$ in this case. The nonrelativistic nature of this thought experiment
should be noted: If the disk has a radius of 5 cm, then $v/c\approx 10^{-6}$ at the rim
of the disk. It is important to emphasize that the nonlocal theory involves the
properties of pure vacuum, whereas in an actual experiment one works with rotating
devices whose characteristics must therefore be known to high accuracy. An interesting
discussion of this point in the case of the emission of radiation by a rotating atomic
system is contained in \cite{15}. Nevertheless, one can hope that future experiments may
achieve levels of accuracy that would make it possible to test predictions (i) and
(ii). In the absence of relevant experimental data of high accuracy, however, it is
useful to employ Bohr's correspondence principle and determine whether quantum
mechanical results are in {\it qualitative} agreement with the predictions of the
classical nonlocal theory. This is the subject of the sections that follow. 

Specifically, in sections~\ref{s3} and \ref{s4} we imagine the rotating observer to be
an electron in a circular Rydberg state of the hydrogen atom. We then consider a
circularly polarized photon that is incident along the normal to the orbital plane. To
study the helicity dependence of the intensity of the radiation field experienced by the
electron, we calculate the dependence of the ionization cross section of the electron
upon the helicity of the incident radiation. To this end, we employ nonrelativistic
quantum mechanics and ignore the spin of the electron. In section~\ref{s5}, we imagine
the rotating observer in (ii) to be an electron in a circular ``orbit" around a uniform
magnetic field. Classically, the circular electron orbit has rotation frequency $\Omega_
c$, which is the cyclotron frequency. We study the helicity dependence of the dipole
transition rates involving a normally incident photon of frequency $\omega =\Omega_c$. 

The main objective of these calculations is to learn via the correspondence principle
what quantum mechanics can teach us about the physics of accelerated systems~\cite{16}.
The results of these studies, based on nonrelativistic quantum mechanics, {\it
qualitatively} bear out, in the correspondence limit, the consequences of the nonlocal
theory of uniformly rotating observers.

It is important to note that, due to the nature of the subject matter, the notation
employed in the following sections is independent of sections~\ref{s1} and \ref{s2},
except when otherwise indicated.

\section{\label{s3}Hypothesis of Locality in Quantum Mechanics}

The hypothesis of locality is necessary for the extension of quantum mechanics to
noninertial frames of reference \cite{17}. Moreover, the impulse approximation scheme in
quantum scattering theory, first discussed by Fermi~\cite{18,19}, turns out to be an
application of the hypothesis of locality to quantum particles. To illustrate this
point, the impulse approximation is employed in this section to show that the cross
section for the ionization of the circular states of atomic hydrogen by a normally
incident circularly polarized plane wave is independent of the helicity of the radiation
in complete correspondence with the standard classical theory (cf. section~\ref{s2}).

The hypothesis of locality involves the replacement of an accelerated system, e.g.
a bound electron, by a free system that is otherwise the same. For instance, in the
ionization of the bound electron by an incident photon, this (impulse) approximation is
valid if the photon energy is much larger than the binding energy of the electron. That
is, during the interaction, the momentum of the electron does not change appreciably
because of its binding force; therefore, the electron may be treated as a free particle. A
free particle of momentum ${\bf p}=\hbar {\bf q}$ has a wave function proportional to
$\exp (i{\bf q}\cdot {\bf x})$; therefore, we must express the normalized wave function
$\Psi ({\bf x},t)=\psi ({\bf x})\exp (-iEt/\hbar)$ of the bound electron in terms of the
Fourier integral
\begin{equation}\label{eq18} \psi ({\bf x})=\frac{1}{(2\pi)^{3/2}}\int d^3q\;\;\hat{\psi}
({\bf q})e^{i{\bf q}\cdot {\bf x}},\end{equation}
where $\hat{\psi} ({\bf q})$ is in effect the momentum-space wave function given by
\begin{equation}\label{eq19} \hat{\psi} ({\bf q})=\frac{1}{(2\pi)^{ 3/2}} \int d^3 x\;\;\psi
({\bf x})e^{-i{\bf q}\cdot {\bf x}}.\end{equation}

To illustrate this application of the hypothesis of locality, we consider the circular
states of the electron in the hydrogen atom and assume that ionization occurs due to the
absorption of a perpendicularly incident photon of frequency $\omega$ such that $Mc^2\gg
\hbar \omega \gg |E_n|$, where $M$ is the mass of the electron and $|E_n|$ is the
electron binding energy. The electron in the final state is free, i.e. we neglect the
Coulomb interaction. The impulse approximation is valid so long as during the
interaction the net impulse due to the Coulomb force can be neglected.

The bound states of the electron in the hydrogen atom are given by the normalized wave
functions
\begin{equation}\label{eq20} \Psi _{n\ell m}(r,\vartheta ,\varphi ,t)=R_{n\ell}(r)Y_{\ell
m}(\vartheta ,\varphi ) e ^{-\frac{iE_nt}{\hbar}}.\end{equation}
The circular states with $n>1$, $\ell =n-1$ and $m=\pm \ell$ correspond to classical
circular orbits in the $(x,y)$ plane. The normalized radial part of the wave function
for $\ell=n-1$ is
\begin{equation}\label{eq21} R_{n\; n-1}(r)=\left( \frac{2}{a_0n}\right)^{3/2}
\frac{1}{\sqrt{(2n)!}} \left( \frac{2r}{a_0n}\right)
^{n-1}e^{-\frac{r}{a_0n}},\end{equation}
where $a_0=\hbar^2/(Me^2)$ is the Bohr radius, $E_n=-e^2/(2r_n)$, $r_n=a_0n^2$ and $-e$ is
the
charge of the electron. We assume for the sake of simplicity that the proton is in effect
fixed at the origin of our spherical polar coordinate system. It follows from
Eq.~\eqref{eq21} that $\langle r\rangle_n=n\left(n+\frac{1}{2}\right) a_0$, so that for
$n\gg 1$, $\langle r\rangle_n\to r_n$.

The circular states of atomic hydrogen have been the subject of extensive experimental
studies, especially in the case of Rydberg atoms (see~\cite{20} and \cite{21} and
references therein). In connection with the ionization of atoms in the correspondence
regime, it is interesting to note that experimental and theoretical studies have been
carried out regarding the ionization of Rydberg atoms by circularly polarized microwave
radiation (see~\cite{22}, \cite{23} and references therein).

Let us now assume an initial counterclockwise circular state with $m=n-1$ and consider
an incident electromagnetic radiation field given by the vector potential
$\bm{\mathcal{A}}$ such that $\bm{\nabla}\cdot \bm{\mathcal{A}}=0$. In this Coulomb gauge,
${\bf p}$ and $\bm{\mathcal{A}}$ commute and the interaction Hamiltonian can be written
as
\begin{equation}\label{eq22} \mathcal{H}_{\text{int}}=\frac{e}{Mc}\bm{\mathcal{A}}\cdot {\bf
p} +\frac{e^2\mathcal{A}^2}{2Mc^2}.\end{equation}
The vector potential may be expressed as
\begin{equation}\label{eq23} \bm{\mathcal{A}}=\sum_{{\bf k}\; \epsilon}\sqrt{\frac{2\pi
\hbar c^2}{\omega V}} ({\bf e}_{{\bf k}\epsilon }a_{{\bf k}\epsilon} e^{-iwt+i{\bf
k}\cdot {\bf x}} +{\bf e}^\ast_{{\bf k}\epsilon} a^\dag_{{\bf k}\epsilon}e^{iwt-i{\bf
k}\cdot {\bf x}}),\end{equation}
where $V$ is the volume of space within a large cube, $\epsilon$ is either plus or minus
and $\omega =ck$. Here, the circular polarization basis for a photon of wave vector
${\bf k}$ is denoted by ${\bf e}_{{\bf k}\epsilon}$; these are so defined that for a
photon propagating along the positive $z$ direction ${\bf e}_{{\bf k}\pm }\to {\bf
e}_\pm$ given in section~\ref{s2}.

We are interested in the ionization cross section due to the absorption of a photon
incident along the $z$ axis. The initial $(|i\rangle )$ and final
$(|f\rangle)$ states are unperturbed
energy eigenstates; indeed, each is a product of the electron state and the photon
state. It follows from the standard first-order time-dependent perturbation theory that
the rate of transition from an initial state to a final state in which the electron is
free is given by
\begin{equation}\label{eq24} dW=\frac{2\pi}{\hbar} |\mathcal{H}_{fi}|^2\delta
(E_f-E_i)\rho _fdE_f,\end{equation}
where $\mathcal{H}_{fi}=\langle f|\mathcal{H}_{\text{int}} |i\rangle $, $\rho_f$ is the
density of final states, i.e. the number of final states per unit energy. The flux of
the incident photon is $c/V$; therefore, the differential cross section for this process
is $d\sigma =dW/(c/V)$. Using energy conservation
\begin{equation}\label{eq25} E_n+\hbar \omega =\frac{\hbar^2{k'}^2}{2M},\end{equation}
where $E_i=E_n+\hbar \omega$ and $E_f=\hbar^2{k'}^2/(2M)$, and the expression for
$\rho_f$
\begin{equation}\rho_f=\frac{V}{(2\pi)^3}\frac{Mk'}{\hbar^2}d\Omega ',\end{equation}
where $dk'_xdk'_y dk'_z={k'}^2dk'd\Omega '$ and $\Omega '$ is the solid angle associated
with the final momentum of the electron $(\hbar {\bf k}')$, we find
\begin{equation}\label{eq27} \frac{d\sigma}{d\Omega '}
=\frac{Mk'V^2}{4\pi^2c\hbar^3}|H_{fi}|^2.\end{equation}
Here, $H_{fi}$ is the reduced matrix element connecting only the electronic states for
which the interaction Hamiltonian reduces to
\begin{equation}\label{eq28} H_{\text{int}}=\frac{e}{M}\sqrt{\frac{2\pi \hbar}{\omega V}}
e^{i{\bf k}\cdot {\bf x}}\; {\bf e}_\pm \cdot {\bf p},\end{equation}
where we have assumed that $e|\bm{\mathcal{A}}|/c$ is much smaller than the electron
momentum, so that the term proportional to $\mathcal{A}^2$ in Eq.~\eqref{eq22} may be
neglected. Thus
\begin{equation}\label{eq29}
H_{fi}=\int d^3x\;\;\psi^\ast_fH_{\text{int}}\psi_i,\end{equation}
where $\psi_f=V^{-1/2}\exp (i{\bf k}'\cdot {\bf x})$ and
\begin{equation}\label{eq30} \psi_i=\frac{1}{(2\pi)^{3/2}}\int d^3q\;\;\hat{\psi}_n ({\bf
q})e^{i{\bf q}\cdot {\bf x}}.
\end{equation}
We find that
\begin{equation}\label{eq31} H_{fi}=\frac{4\pi^2e\hbar^{3/2}}{MV\omega^{1/2}} {\bf e}_\pm \cdot ({\bf k}'-{\bf
k})\;\;\hat{\psi}_n ({\bf k}'-{\bf k}),\end{equation}
where ${\bf e}_\pm \cdot {\bf k}=0$ by assumption and ${\bf e}_\pm \cdot {\bf k}'=\mp
(k'_x\pm ik'_y)/\sqrt{2}$. Thus
\begin{equation}\label{eq32} \frac{d\sigma}{d\Omega '}=\frac{2\pi^2\alpha \hbar k'}{M\omega}
({k'}^2_x+{k'}^2_y)|\hat{\psi}_n({\bf k}'-{\bf k})|^2,\end{equation}
where $\alpha =e^2/(\hbar c)$ is the fine-structure constant. It is already clear from
Eq.~\eqref{eq32} that there is
no helicity-dependent photoionization in the impulse approximation. The general nature of
this result should be
emphasized, since it is independent of the nature of the initial state.

For comparison purposes, it is useful to compute $d\sigma /d\Omega '$ explicitly. To this end, we note that (see
Appendix~\ref{aB})
\begin{equation}\label{eq33}\hat{\psi}_n({\bf q})=\frac{1}{2\pi} \sqrt{\frac{2}{a_0}}(2a_0n)^{n+1}
\frac{(iqe^{i\phi_q}\sin \theta _q)^{n-1}}{(1+a^2_0 n^2q^2)^{n+1}},\end{equation}
where $\hat{\bf q}= (\theta_q,\phi_q)$. From $\hat{\bf k}'=(\theta ,\phi)$ and ${\bf q}={\bf k}'-{\bf k}$, we find
that
\begin{equation}\label{eq34} q\sin \theta_q=k'\sin \theta ,\quad q\cos \theta_q=k'\cos \theta -k,\quad \phi_q=\phi
.\end{equation}
Thus Eq.~\eqref{eq32} may be written as
\begin{equation}\label{eq35} \frac{d\sigma}{d\Omega '}=4\alpha a^2_0n^3 \frac{|E_n|}{\hbar \omega} (2a_0nk')^{2n+1}
\frac{\sin^{2n}\theta }{(1+a^2_0n^2q^2)^{2n+2}},\end{equation}
where $q^2=k^2+{k'}^2-2kk'\cos \theta$. In the impulse approximation, $Mc^2\gg \hbar \omega \gg
|E_n|$; therefore,
$k/k'\approx v'/(2c)\ll 1$,
\begin{equation}\label{eq36} a_0nk'\approx \left( \frac{\hbar \omega}{|E_n|}\right)^{\frac{1}{2}}, \quad 
1+a^2_0n^2q^2\approx \frac{\hbar \omega}{|E_n|} \left(1-\frac{v'}{c}\cos \theta \right),\end{equation}
where $v'=\hbar k'/M$. Hence, we find that
\begin{equation}\label{eq37} \frac{d\sigma }{d\Omega '}\approx 2^{2n+3}\alpha a^2_0 n^3\left( \frac{|E_n|}{\hbar
\omega}\right)^{n+\frac{5}{2}} \frac{\sin ^{2n}\theta}{\left( 1-\frac{v'}{c}\cos \theta \right)^{2n+2}}.\end{equation}
The total ionization cross section is then (see Appendix~\ref{aB})
\begin{equation}\label{eq38}\sigma \approx 2^{4n+5} \pi \alpha a^2_0n^3 \frac{(n!)^2}{(2n+1)!} \left(
\frac{|E_n|}{\hbar \omega }\right) ^{n+\frac{5}{2}},\end{equation}
which is valid to $O({v'}^2/c^2)$, since the term linear in $v'/c$ vanishes. Though this formula has been derived for
circular states $(n>1)$, it applies equally well to the $n=1$ spherically symmetric ground state of the hydrogen
atom~\cite{24}.

The calculation of $\sigma$ has been based on the impulse approximation, which means
that during the electron-photon interaction, the change in the momentum of the electron
due to the Coulomb binding force has been neglected. However, the ionization cross
section is expected to become dependent upon the helicity of the incident radiation when
the Coulomb interaction is fully taken into account.

\section{\label{s4}Helicity-Dependent Photoeffect}

The standard treatment of the photoeffect is contained in \cite{25,26}. The purpose of this section is to go beyond
the impulse approximation of section~\ref{s3} and show that the inclusion of the Coulomb interaction in the final
state leads to the helicity-dependent photoeffect that is in qualitative agreement with the
nonlocal prediction (i) of section~\ref{s2}.

As in section~\ref{s3}, we assume that a hydrogen atom is initially in a circular state of energy $E_n$. A photon with
wave vector ${\bf k}=(\omega /c)\hat{\bf z}$ is normally incident on the electron orbit such that
\begin{equation}\label{eq39} Mc^2\gg \hbar \omega >|E_n|,\end{equation}
leading to ionization. In dealing with the reduced matrix element $H_{fi}$, we take the Coulomb interaction into
account in the final state wave function. But to simplify matters, we use the electric
dipole approximation, i.e. $\exp (i{\bf k}\cdot {\bf x})\approx 1$. It follows that $\omega
r_n\ll c$, where $r_n=a_0n^2$, and hence
\begin{equation}\label{eq40} \frac{1}{2}\alpha \ll \frac{|E_n|}{\hbar \omega} <1.\end{equation}

Let $\psi _n$ be the initial state and $\psi_C$ be the final Coulomb state, then
\begin{equation}\label{eq41} H^\pm_{fi}=\frac{e}{M}\sqrt{\frac{2\pi \hbar}{\omega V}} \int d^3
x\;\; \psi^\ast_C ({\bf
e}_\pm \cdot {\bf p}) \psi_n.\end{equation}
We can replace the momentum  operator ${\bf
p}$ with $(-iM/\hbar)[{\bf
x},H_0]$, where $H_0$ is the unperturbed hydrogen
Hamiltonian. It follows from energy conservation, Eq.~\eqref{eq25}, that
\begin{equation}\label{eq42} H^\pm _{fi}=ie \sqrt{\frac{2\pi \hbar \omega}{V}}\int
d^3x\;\;\psi^\ast _C ({\bf e}_\pm \cdot
{\bf x}) \psi_n.\end{equation}
We note that 
\begin{equation}\label{eq43} {\bf e}_\pm \cdot {\bf x}=\sqrt{\frac{4\pi}{3}}\; r\;
Y_{1\;\pm 1}
(\vartheta,\varphi
).\end{equation}
Moreover, $\psi^\ast_C$ can be expressed as (see~\cite{19}, p. 470)
\begin{equation}
\label{eq44} \psi^\ast_{C}=\frac{4\pi}{\sqrt{V}} \sum_{\ell m} i^{-\ell}C_\ell
(k';r)Y_{\ell m}(\theta ,\phi )Y^\ast _{\ell m}(\vartheta ,\varphi),
\end{equation}
where
\begin{equation}\label{eq45} C_\ell (k';r)=\frac{(2k'r)^\ell e^{\frac{1}{2}\pi \gamma
+ik'r}}{(2\ell +1)!} \Gamma (\ell +1-i\gamma )F(\ell +1-i\gamma ,2\ell
+2,-2ik'r
).\end{equation}
Here, $\gamma^{-1}=k'a_0$ and $F$ is a confluent hypergeometric function. Let us note
that if $C_\ell (k';r)$ is replaced by $j_\ell (k'r)$ in Eq.~\eqref{eq44}, then $\psi
^\ast_C\to \psi^\ast_f$ of section~\ref{s3}. Equation~\eqref{eq42} can be expressed as
\begin{equation}\label{eq46} H^\pm _{fi}=\frac{8\pi ^2ie}{V}\sqrt{\frac{2\hbar
\omega}{3}} \sum_{\ell m}i^{-\ell }I^\pm _{\ell m}\mathcal{C}_\ell Y_{\ell m}(\theta
,\phi),\end{equation}
where
\begin{eqnarray}\label{eq47} I^\pm_{\ell m}&=&\int d\Omega\;\; Y^\ast _{\ell m} (\Omega
)Y_{1\;\pm 1} (\Omega )Y_{n-1\; n-1}(\Omega),\\
\label{eq48} \mathcal{C}_\ell &=& \int^\infty_0 dr\;\;
r^3R_{n\; n-1}(r)C_\ell
(k';r).\end{eqnarray}
It follows from a standard result (see~\cite{19}, p. 290) that
\begin{equation}\label{eq49} I^\pm_{\ell m}=\sqrt{\frac{3(2n-1)}{4\pi (2\ell+1)}}
\langle \ell ',\ell ',1,\pm 1\mid\ell ,m\rangle \langle \ell ',0 ,1,0\mid \ell
,0\rangle,\end{equation}
where $\ell '=n-1$. From the general properties of the Clebsch-Gordan coefficients, it
is clear that $I^+_{\ell m}$ can be nonzero only for $\ell =m=n$, while $I^-_{\ell m}$
can be nonzero only for $m=n-2$ and $\ell =n$, $n-1$, $n-2$. Using the table on p. 220
and formula (34.40) on p. 290 of \cite{19}, one finds that
\begin{eqnarray}\label{eq50}
I^+_{n\; n}&=&
\sqrt{\frac{3n}{4\pi (2n+1)}} ,\quad 
I^-_{n\;n-2}=\sqrt{\frac{3}{4\pi (4n^2-1)}},\\
\label{eq51} I^-_{n-1\; n-2}&=&0,\quad  I^-_{n-2\; n-2}=-\sqrt{\frac{3(n-1)}{4\pi
(2n-1)}} .\end{eqnarray}
Therefore, 
\begin{eqnarray}\label{eq52} H^+_{fi}&=& -\frac{4\pi e}{Vi^{n+1}} \sqrt{\frac{2\pi n\hbar
\omega}{2n+1}} \mathcal{C}_nY_{nn} (\theta ,\phi ),\\
\label{eq53} H^-_{fi}&=&-\frac{4\pi e}{Vi^{n+1}}\sqrt{\frac{2\pi (n-1)\hbar \omega}{2n-1}}
\Bigl[ \mathcal{C}_{n-2} Y_{n-2\; n-2}(\theta ,\phi )\nonumber\\
&+&\frac{1}{\sqrt{(n-1)(2n+1)}}
\mathcal{C}_nY_{n\; n-2}(\theta ,\phi )\Bigr].\end{eqnarray}

Computing the total cross section
\begin{equation}\label{eq54} \sigma_\pm =\frac{Mk'V^2}{4\pi ^2c\hbar ^3}\int
|H^\pm_{fi}|^2 d\Omega ',\end{equation}
we find that
\begin{eqnarray}\label{eq55} \sigma_+&=& 8\pi \frac{kk'}{a_0} \frac{n}{2n+1}
|\mathcal{C}_n|^2,\\
\label{eq56} \sigma_-&=& 8\pi \frac{kk'}{a_0} \left[ \frac{n-1}{2n-1}
|\mathcal{C}_{n-2}|^2+\frac{1}{4n^2-1} |\mathcal{C}_n|^2\right].\end{eqnarray}
The quantities $\mathcal{C}_n$ and $\mathcal{C}_{n-2}$ can be calculated using the
results given in Appendix~\ref{aB}. Hence, 
\begin{equation}\label{eq57} \mathcal{C}_n=\sqrt{8}(a_0n)^{5/2} \left( \frac{4n}{\gamma
}\right) ^n \frac{\Gamma (n+1-i\gamma )}{\sqrt{(2n)!}} \left( 1+\frac{n^2}{\gamma
^2}\right) ^{-n-2}e^{\frac{1}{2}n\gamma -2\gamma \cot ^{-1}\left( \frac{\gamma
}{n}\right)} .\end{equation}
Similarly, we find that
\begin{equation}\label{eq58} \frac{\mathcal{C}_{n-2}}{\mathcal{C}_n}=\frac{1}{2n}\;\;
\frac{n+i\gamma}{n-1-i\gamma }.\end{equation}
It follows from these results that
\begin{equation}\label{eq59}
\frac{\sigma_-}{\sigma_+}=\frac{3n^2(n-1)+(3n+1)\gamma^2}{4n^3[(n-1)^2+\gamma^2]},\end{equation}
which is valid for $n=1,2,3,\ldots $. The $n=1$ ground state of hydrogen is spherically
symmetric; therefore, $\sigma_+=\sigma_-$ in agreement with Eq.~\eqref{eq59}. For circular
states with $n>1$, Eq.~\eqref{eq59} implies that $\sigma _-<\sigma_+$ in correspondence
with the nonlocal theory of section~\ref{s2}.

To bring out this qualitative agreement more explicitly, we recall that for a Bohr orbit
of speed $v_n=c\;\; \alpha /n$ and radius $r_n$, one can define a Bohr frequency $\Omega_n$
given by $v_n=r_n\Omega_n$; it is then simple to show that $\Omega_n=2|E_n|/(\hbar n)$.
Let us consider the ratio $\eta :=\Omega _n/\omega$ for a circular state. Then, from
Eq.~\eqref{eq25} and $\gamma^{-1}=k'a_0$ we find that 
\begin{equation}\label{eq60} \gamma^2=\frac{n^3\eta}{2-n\eta }.\end{equation}
Substituting this relation in the expression for $\sigma_-/\sigma_+$ results in
\begin{equation}\label{eq61} \frac{\sigma_-}{\sigma_+}=\frac{3(n-1) +2n\eta
}{2n[2(n-1)^2+n(2n-1)\eta ]},\end{equation}
so that for a given circular state $n$, $\sigma _-/\sigma_+$ only depends on $\eta
=\Omega_n/\omega$ in agreement with the classical nonlocal theory. We note that
Eq.~\eqref{eq40} can be written in terms of $\eta$ as
\begin{equation}\label{eq62} \frac{\alpha }{n}\ll \eta < \frac{2}{n};\end{equation}
therefore, for $n\geq 2$ this approach can be qualitatively compared with the nonlocal
theory, cf. Eq.~\eqref{eq16}.

The treatment of ionization presented here can be used near threshold $(\gamma \to
\infty)$ as well as for $\hbar \omega $ in the intermediate energy regime given by
$Mc^2\gg \hbar \omega \gg |E_n|$, where $\gamma \ll 1$. The threshold behavior has been
discussed, for instance, in \cite{19} and \cite{26}; therefore, we concentrate on the
latter case $(\gamma \ll 1)$ that was treated in the previous section using the impulse
approximation. In general, it is possible to express $n/\gamma$ as
\begin{equation}\label{eq63} \frac{n}{\gamma }=\sqrt{\frac{\hbar
\omega}{|E_n|}-1},\end{equation}
so that for $\gamma \ll 1$, $\hbar \omega\gg |E_n|$ and
\begin{equation}\label{eq64} \frac{n}{\gamma }\approx \left( \frac{\hbar
\omega}{|E_n|}\right)^{1/2}\gg 1.\end{equation}
It follows that $d\sigma_+ /d\Omega '$ reduces in this case to the result of the impulse
approximation $d\sigma /d\Omega '$ given in Eq.~\eqref{eq37}, i.e. for $\gamma \ll 1$,
\begin{equation}\label{eq65} \frac{d\sigma_+}{d\Omega '}\approx \frac{d\sigma}{d\Omega
'}.\end{equation}
For $n\geq 2$, however, $d\sigma _-/d\Omega '$ does not approach $d\sigma /d\Omega '$
for $\gamma \ll 1$; in fact, $\sigma_-/\sigma_+\approx 3/[4n(n-1)]$ in this case. This
helicity dependence is a noteworthy aspect of the Coulomb interaction. It would be
interesting to investigate experimentally this helicity dependence of the photoeffect
for $n\geq 2$.

\section{\label{s5}Resonant Absorption}

Let us now consider the motion of an electron in a uniform magnetic field ${\bf
B}=B\hat{\bf z}$. We are interested in circular orbits about the magnetic lines of
force; therefore, we consider the solution of Schr\"odinger's equation in cylindrical
coordinates $(\rho ,\varphi ,z)$, i.e.
\begin{equation}\label{eq66} \frac{1}{2M} \left({\bf p} +\frac{e}{c}{\bf A}\right)^2\Psi
=i\hbar \frac{\partial \Psi}{\partial t},\end{equation}
where ${\bf A}=\frac{1}{2}B\rho \hat{\bm{\varphi}}$. It is useful to introduce a
magnetic length $\rho_0$, $\rho^2_0=\hbar c/(eB)$. The nonrelativistic treatment is
valid so long as the Compton wavelength of the electron is much smaller than the
magnetic radius $\rho_0$; this requirement can be satisfied for $B\ll M^2c^3/(e\hbar )$.

In terms of a new dimensionless radial variable $\xi$, $\xi =\rho ^2/(2\rho ^2_0)$, the solutions of
Eq.~\eqref{eq66} can be expressed in terms of the confluent hypergeometric functions~\cite{27}. The acceptable
solutions of the Schr\"odinger equation~\eqref{eq66} are of the form
\begin{equation}\label{eq67} \Psi =C_0e^{im\varphi +i\frac{p_z}{\hbar} z-i\frac{E}{\hbar} t}\chi (\xi ),\end{equation}
where $C_0$ is a normalization constant, $m$ is the azimuthal quantum number and
\begin{equation}\label{eq68} \chi
(\xi )=\xi ^{\frac{1}{2}|m|}e^{-\frac{1}{2}\xi} L^{|m|}_{n_\rho} (\xi).\end{equation} Here, $n_\rho =0,1,2,\dots $,
is the radial quantum number and $L^{|m|}_{n_\rho}$ is an associated Laguerre polynomial (see
Appendix~\ref{aC}). The
energy of the electron is given by
\begin{equation}\label{eq69} E=\frac{p^2_z}{2M}+\hbar \Omega_c\left( n+\frac{1}{2}\right),\end{equation}
where $\Omega_c=eB/(Mc)$ is the cyclotron frequency, $\hbar \Omega_c\ll Mc^2$, and
\begin{equation}\label{eq70} n=n_\rho +\frac{m+|m|}{2}.\end{equation}

To discuss the correspondence limit, it is convenient to define the following three Hermitian operators:
\begin{equation}\label{eq71} H^c=\frac{p^2_z}{2M}+\frac{1}{2}M\Omega^2_c\rho^2,\quad
J^c=M\Omega_c\rho^2\end{equation} and $\bm{\ell} =M{\bf r}\times {\bf v}$, where ${\bf v}$ is defined by ${\bf
p}=M{\bf v} -e{\bf A}/c$. Let us note that $H^c$ corresponds to the classical energy of the electron, $J^c$
corresponds to its angular momentum about the $z$ axis and $\bm{\ell}$ is the operator of
the classical orbital
angular momentum. Thus $\bm{\ell }={\bf L}+(e/c){\bf r}\times {\bf A}$ and $\ell_z=L_z+eB\rho^2/(2c)$. We find that
in the eigenstate given by Eq.~\eqref{eq67}, $\langle \xi\rangle =2n_\rho +|m|+1$, so that
\begin{eqnarray}\label{eq72} \langle H^c\rangle &=& \frac{p^2_z}{2M}+\hbar \Omega_c(2n_\rho +|m|+1),\\
\label{eq73} \langle J^c\rangle &=& 2\hbar (2n_\rho +|m|+1),\\
\label{eq74} \langle \ell_z\rangle &=& 2\hbar \left( n_\rho +\frac{m+|m|}{2}+\frac{1}{2}\right).\end{eqnarray}
Clearly, $H^c=\frac{p^2_z}{2M}+\frac{1}{2}\Omega_cJ^c$ and $E=\frac{p^2_z}{2M}+\frac{1}{2} \Omega_c\langle \ell
_z\rangle$. We expect that in the correspondence limit $E\sim \langle H^c\rangle$ and hence $\langle J^c\rangle
\sim\langle \ell_z\rangle $, so that the wave function with $m\gg 1$ and $m\gg n_\rho$
would correspond to classical
orbits based on the comparison between Eqs.~\eqref{eq69} and \eqref{eq72}.

We are interested in the transition of the electron to a state of higher energy as a result of the resonant absorption
of a photon of frequency $\omega =\Omega_c$ that is normally incident along  the $z$ direction. We therefore assume
that the electron is initially in a circular ``orbit" with energy $\mathcal{E}_i$, $p_z=0$ and $n_i=n_\rho +m_i$,
where $m_i\gg 1$ and $m_i\gg n_\rho$. Conservation of energy and momentum imply that the excited state should have
energy $\mathcal{E}_f=\mathcal{E}_i+\hbar \omega$ and momentum $p_z=\hbar \omega /c$ with $\omega =\Omega_c$. Thus the
initial and final principal quantum numbers are related by $n_f-n_i=1-\hbar \Omega_c/(2Mc^2)$; however, we neglect
$\hbar \Omega _c/(Mc^2)\ll 1$ in our nonrelativistic approximation scheme and set $n_f$ equal to $n_i+1$.

The interaction Hamiltonian is given by
\begin{equation}\label{eq75} \mathcal{H}_{\text{int}}=\frac{e}{Mc}\bm{\mathcal{A}}\cdot \left({\bf p}+\frac{e}{c}{\bf
A}\right) + \frac{e^2\mathcal{A}^2}{2Mc^2},\end{equation}
where $\bm{\mathcal{A}}$ is given by Eq.~\eqref{eq23}. Assuming that the incident radiation is sufficiently weak,
i.e, $e|\bm{\mathcal{A}}|/c$ is very small compared to the electron momentum, we neglect the term proportional to
$\mathcal{A}^2$ in Eq.~\eqref{eq75}. The transition probability for the {\it ideal} case of resonant absorption can be
simply worked out using first-order time-dependent perturbation theory~\cite{27}, and the result is
\begin{equation}\label{eq76} P=\frac{1}{\hbar^2} |\langle f|\mathcal{H}_{\text{int}} |i\rangle |^2t^2;\end{equation}
clearly, the validity of Eq.~\eqref{eq76} is limited in time. Here $|i\rangle $ and $|f\rangle$ are unperturbed energy
eigenstates; in fact, each is a product of the electron state and the photon state. To simplify matters, line
broadening is totally neglected here; in particular, the unperturbed states are assumed to have infinite lifetimes.

Concentrating on the matrix element $\langle f|\mathcal{H}_{\text{int} }|i\rangle$, we note that $\mathcal{H}_{\text{int} }$ reduces
to
\begin{equation}\label{eq77} H_{\text{int} } =\frac{e}{M}\sqrt{\frac{2\pi
\hbar}{\Omega_cV}}\;\; {\bf e}_\pm \cdot \left({\bf
p}+\frac{e}{c}{\bf A}\right)\end{equation}
that acts only on the electronic states. It can thus be expressed as
\begin{equation}\label{eq78} H_{\text{int} } =\pm i\hbar e\sqrt{\frac{\pi}{2MV}} e^{\pm i\varphi} \xi^{\frac{1}{2}} 
\left( 2\frac{\partial}{\partial \xi} \pm \frac{i}{\xi } \frac{\partial }{\partial \varphi} \mp
1\right).\end{equation}
It follows from the conservation of angular momentum that we must have $m_f=m_i\pm 1$, since a photon of helicity $\pm
1$ carries an angular momentum of $\pm \hbar$ along its direction of motion. Writing $n_f=n'_\rho +m_f$ and $n_i=n_\rho
+m_i$, we find that $n'_\rho =n_\rho $ in the positive helicity case and $n'_\rho =n_\rho +2$ in the negative
helicity case. Thus the computation of the matrix element $\langle
f|\mathcal{H}_{\text{int}}|i\rangle$ reduces to the
evaluation of the integrals \begin{equation}\label{eq79} \mathcal{I}_\pm =\int^\infty_0\chi^\ast
_f\;\;\xi^{\frac{1}{2}}
\left( 2\frac{\partial }{\partial \xi} \mp \frac{m_i}{\xi } \mp 1\right) \chi _i \;\; d\xi
,\end{equation}
since $\rho d\rho d\varphi =\rho ^2_0 d\xi d\varphi$. Here, $\chi_i$ and $\chi_f$ are given by
Eq.~\eqref{eq68} for the initial and final states. It turns out that (see Appendix~\ref{aC})
\begin{equation}\label{eq80} \mathcal{I}_+ =-2\frac{(n_\rho +m_i+1)!}{n_\rho !} ,\quad
\mathcal{I}_-=0.\end{equation}

The correspondence principle connects the square of the amplitude of the classical field measured by the accelerated
observer with the probability of transition. In the positive-helicity case, both of these
functions increase
quadratically with time, while in the negative-helicity case the classical field is periodic in time and averages to
zero in agreement with the fact that the transition probability vanishes in quantum mechanics. We conclude that the
qualitative results of first-order perturbation theory for resonant absorption are consistent with the nonlocal
electrodynamics of uniformly rotating observers.

\section{\label{s6}Discussion}

The nonlocal theory of accelerated observers is an attempt at the simplest physical theory that has a consistent
mathematical structure and goes beyond the standard theory that is based on the hypothesis of locality, namely, the
assumption that an observer's acceleration is irrelevant at each instant for measurement purposes, so that the
accelerated observer is equivalent to a hypothetical momentarily comoving inertial observer. The nonlocal theory
involves an averaging procedure over the past worldline of an accelerated observer; the corresponding weighting function
is a kernel that represents the memory of past acceleration. The consequences of this theory have been worked out in
the present paper for a uniformly rotating observer. These are compared with the helicity dependence of the rates of
ionization of circular states of atomic hydrogen as well as the helicity dependence of the transition probabilities
for electrons in circular ``orbits" about a uniform magnetic field. The nonlocal results agree better with quantum
mechanics in the correspondence limit than the standard relativistic theory of accelerated observers based on the
locality hypothesis.

\appendix

\section{}\label{aA}

A simple way to derive Eq.~\eqref{eq15} is via the following general result: Substituting Eq.~\eqref{eq6} for the
kernel in Eq.~\eqref{eq5}, then using $\hat{F}=\Lambda F$ and integration by parts, we obtain for $\tau >\tau_0$
\begin{equation}\label{eqA1} \hat{\mathcal{F}}(\tau ) =\hat{F}(\tau _0)+\int^\tau_{\tau_0} \Lambda (\tau
')\frac{dF}{d\tau '}d\tau '.\end{equation}
Let us now specialize to the case of a monochromatic radiation field such that $F$ varies with proper time as $\exp
(-i\gamma \omega \tau)$; then,
\begin{equation}\label{eqA2} \hat{\mathcal{F}}(\tau )=\hat{F}(\tau _0)-i\gamma \omega \int^\tau_{\tau _0}\hat{F}(\tau
')d\tau '.\end{equation}
Suppose that, as in Eq.~\eqref{eq13}, $\hat{F}(\tau )$ varies with proper time as $\exp (-i\hat{\omega }\tau )$; then,
Eq.~\eqref{eqA2} implies that
\begin{equation}\label{eqA3} \hat{\mathcal{F}} (\tau )=\hat{F}(\tau _0) \left[ 1+\gamma \omega
\frac{e^{-i\hat{\omega}(\tau -\tau _0)}-1}{\hat{\omega}}\right].\end{equation}
Substituting Eq.~\eqref{eq13} for $\hat{F}$ and setting $\tau_0=0$ in Eq.~\eqref{eqA3}, we recover Eq.~\eqref{eq15}.

\section{}\label{aB} 

In computing the Fourier integral of Eq.~\eqref{eq20}, the following relations have been used:
\begin{eqnarray} \label{eqB1}  e^{-i{\bf q}\cdot {\bf x}}= 4\pi \sum_{\ell m}(-i)^\ell
j_\ell (qr)Y_{\ell m}(\theta_q, \phi _q)Y^\ast _{\ell m}(\vartheta ,\varphi),\\
\label{eqB2} Y_{\ell \ell}(\hat{\bf q}) =\frac{(-1)^\ell}{2^\ell \ell !}
\sqrt{\frac{(2\ell+1)!}{4\pi}} e^{i\ell
\phi_q}\sin ^\ell \theta _q,\\
\label{eqB3}  \int^\infty_0\rho^{n+1}e^{-\rho} j_{n-1}(\lambda \rho)d\rho =2^nn!
\frac{\lambda ^{n-1}}{(1+\lambda^2)^{n+1}}.
\end{eqnarray}
This integral follows from the second formula in (6.623) on page 712 of \cite{28}, namely,
\begin{equation}\label{eqB4} \int^\infty_0e^{-\alpha x} J_\nu (\beta x)x^{\nu
+1}dx=\frac{2\alpha (2\beta )^\nu
\Gamma \left( \nu+\frac{3}{2}\right)}{\sqrt{\pi}(\alpha^2+\beta^2)^{\nu+\frac{3}{2}}},\end{equation}
where $\text{Re } \nu >-1$ and $\text{Re } \alpha >|\text{Im } \beta |$.
Using
\begin{equation}\label{eqB5}j_\ell(\rho )
=\sqrt{\frac{\pi}{2\rho}}J_{\ell +\frac{1}{2}} (\rho ),\end{equation}
$\alpha =1$, $\beta =\lambda$ and $\nu =n+\frac{1}{2}$, we get Eq.~\eqref{eqB3} with $n\to n+1$.

The evaluation of the total cross section in Eq.~\eqref{eq38} is based on the relation
\begin{equation} \int^\pi_0\sin^{2n+1}\theta\; d\theta =2^{2n+1} \frac{(n!)^2}{(2n+1)!}.\end{equation}

To calculate $\mathcal{C}_n$ and $\mathcal{C}_{n-2}$ in Eqs.~\eqref{eq57} and \eqref{eq58},
one can use the relation (see~\cite{27}, \S f of Mathematical Appendices)
\begin{equation}\label{eqB7} \int^\infty_0e^{-\zeta z}z^\nu F(a,b,Kz)\; dz=(-1)^N\Gamma
(b)\frac{d^N}{d\zeta ^N}[\zeta^{a-b}(\zeta -K)^{-a}],
\end{equation}
where $\nu +1=b+N$, $N=1,2,3,\ldots$, and
\begin{equation}\label{eqB8}\text{Re } (b-a)>0,\quad \text{Re }(b+N)>0,\quad \text{Re } \zeta
>|\text{Re }K|.\end{equation}
To calculate $\mathcal{C}_n$, one can change the integration variable in Eq.~\eqref{eq48} to $r/(a_0n)$ and let $N=1$,
$\zeta =1-in/\gamma$, $\nu=b=2n+2$, $a=n+1-i\gamma$ and $K=-2in/\gamma$. In Eq.~\eqref{eq57}, we have used the
relation
\begin{equation}\label{eqB9} \left( \frac{\gamma +in}{\gamma -in}\right)^{i\gamma}=e^{-2\gamma \cot
^{-1}\left(
\frac{\gamma }{n}\right)};\end{equation}
moreover, $\Gamma (n+1-i\gamma)$ can be computed using $\Gamma (1+z)=z\Gamma (z)$ and
\begin{equation}\label{eqB10} \Gamma (z)\Gamma (-z)=-\frac{\pi}{z\sin \pi z}.\end{equation}
It follows that
\begin{equation}\label{eqB11} |\Gamma (n+1-i\gamma
)|^2=\frac{\pi\gamma}{\sinh \pi
\gamma}\prod^n_{s=1}(s^2+\gamma^2).\end{equation}
Furthermore, regarding the phase of $\Gamma (n+1-i\gamma)$, we note that \cite{26}
\begin{equation}\label{eqB12} \frac{\Gamma (n+1-i\gamma )}{\Gamma (n+1+i\gamma )}\approx 1-2i\gamma \ln \left(
n+\frac{1}{2} \right) -2\gamma ^2\ln ^2\left( n+\frac{1}{2}\right) +\dots \end{equation}
for $\gamma \ll 1$ and $n\gg 1$. More generally, one can use Stirling's series that is an asymptotic expansion given
by
\begin{equation}\label{eqB13} \ln \Gamma (1+z)=\frac{1}{2}\ln 2\pi +\left( z+\frac{1}{2}\right) \ln
z-z+\sum^\infty_{n=1}\frac{B_{2n}}{2n(2n-1)}z^{1-2n} ,\end{equation}
where the $B_{2n}$ are Bernoulli numbers.

For the calculation of $\mathcal{C}_{n-2}$, we change the integration variable the same way
as before and let
$N=3,\zeta =1-in/\gamma$, $\nu =2n$, $a=n-1-i\gamma$, $b=2n-2$ and, as before, $K=-2in/\gamma$. It follows from a long
but straightforward calculation that $\mathcal{C}_{n-2}$ is given by Eq.~\eqref{eq58}.

\section{}\label{aC} 

The associated Laguerre polynomial $L_n^{\;\; k}(x)$ is defined by the generating function
\begin{equation}\label{eqC1} \frac{e^{-\frac{xz}{1-z}}}{(1-z)^{k+1}}=\sum^\infty_{n=0} L_n^{\;\; k}
(x)z^n\end{equation} for $|z|<1$ and $k=0,1,2,\dots$.

In the calculation of $\mathcal{I}_\pm$ in Eq.~\eqref{eq79}, the following  relations were used
\begin{eqnarray}\label{eqC2} \frac{dL_{n+1}^{\;\;k}(x)}{dx}&=& -L_n^{k+1}(x),\\
\label{eqC3} L_n^{\;\;k+1} &=& L^{\;\; k}_n +L_{n-1}^{\;\; k}+\dots +L^{\;\; k}_0,\end{eqnarray}
as well as the orthogonality property
\begin{equation}\label{eqC4} \int^\infty_0e^{-x}x^kL_m^{\;\; k} L_n^{\;\; k} dx=\frac{(n+k)!}{n!}
\delta_{mn}.\end{equation} From Eq.~\eqref{eqC3} one gets the useful relation
\begin{equation}\label{eqC5}L_n^{\;\;k}=L_n^{\;\; k+1}-L_{n-1}^{\;\;\;\;k+1}.\end{equation}


\begin{thebibliography}{xxxxxx}

\bibitem{1} A. Einstein, {\it The Meaning of Relativity} (Princeton University Press, Princeton, NJ, 1955).
\bibitem{2} B. Mashhoon, Phys. Lett. A {\bf 143}, 176 (1990); Phys. Lett. A {\bf 145}, 147 (1990); in {\it Relativity
in Rotating Frames}, edited by G. Rizzi and M.L. Ruggiero (Kluwer Academic Publishers, Dordrecht, 2003), pp. 43-55.
\bibitem{3} B. Mashhoon, Phys. Rev. A {\bf 47}, 4498 (1993).
\bibitem{4} N. Bohr and L. Rosenfeld, K. Dan. Vidensk. Selsk. Mat. Fys. Medd. {\bf 12}, No. 8, (1933); translated in
{\it Quantum
Theory and Measurement}, edited by J.A. Wheeler and W.H. Zurek (Princeton University Press,
Princeton, NJ, 1983); Phys. Rev. {\bf 78}, 794 (1950).
\bibitem{5} V. Volterra, {\it Theory of Functionals and of Integral and Integro-Differential Equations} (Dover, New York, 1959).
\bibitem{6} F.G. Tricomi, {\it Integral Equations} (Interscience, New York, 1957).
\bibitem{7} C. Chicone and B. Mashhoon, Ann. Phys. (Leipzig) {\bf 11}, 309 (2002).
\bibitem{8} C. Chicone and B. Mashhoon, Phys. Lett. A {\bf 298}, 229 (2002).
\bibitem{9} F.W. Hehl and Y.N. Obukhov, {\it Foundations of Classical Electrodynamics} (Birkh\"auser, Boston, 2003).
\bibitem{10} B. Mashhoon, Phys. Rev. A {\bf 70}, 062103 (2004).
\bibitem{11} B. Mashhoon, Phys. Lett. A {\bf 139}, 103 (1989); B. Mashhoon, R. Neutze, M.
Hannam and G.E. Stedman, Phys. lett.
A {\bf 249}, 161 (1998); J.C. Hauck and B. Mashhoon, Ann. Phys. (Leipzig) {\bf 12}, 275
(2003).
\bibitem{12} N. Ashby, Living Rev. Relativity {\bf 6}, 1 (2003).
\bibitem{13} B. Mashhoon, Phys. Lett. A {\bf 306}, 66 (2002); J.D. Anderson and B.
Mashhoon, Phys. Lett. A {\bf 315}, 199 (2003).
\bibitem{14} B. Mashhoon, Gen. Rel. Grav. {\bf 31}, 681 (1999); Class. Quantum Grav. {\bf 17}, 2399 (2000).
\bibitem{15} I. Bialynicki-Birula and Z. Bialynicka-Birula, Phys. Rev. Lett. {\bf 78}, 2539 (1997).
\bibitem{16} I owe this idea to Steven Chu.
\bibitem{17} B. Mashhoon, Phys. Rev. Lett. {\bf 61}, 2639 (1988).
\bibitem{18} E. Fermi, Ric. Sci. {\bf VII-11}, 13 (1936); R.G. Newton, {\it Scattering Theory of Waves and Particles},
2nd ed. (Springer-Verlag, New York, 1982); M.L. Goldberger and K.M. Watson, {\it Collision Theory} (Wiley, New York,
1964).
\bibitem{19} K. Gottfried, {\it Quantum Mechanics} (Benjamin, New York, 1966).
\bibitem{20} R.G. Hulet and D. Kleppner, Phys. Rev. Lett. {\bf 51}, 1430 (1983).
\bibitem{21} R. Lutwak, J. Holley, P.P. Chang, S. Paine, D. Kleppner and T. Ducas, Phys. Rev. A {\bf 56}, 1443 (1997).
\bibitem{22} P. Fu, T.J. Scholz, J.M. Hettema and T.F. Gallagher, Phys. Rev. Lett. {\bf 64}, 511 (1990).
\bibitem{23} A.F. Brunello, T. Uzer and D. Farrelly, Phys. Rev. A {\bf 55}, 3730 (1997);
T. Cheng, J. Liu, S. Chen and H. Guo, Phys. Lett. A {\bf 265}, 384 (2000).
\bibitem{24} In connection with $\sigma$ for $n=1$, there is a misprint in Eq. (58.14) of \cite{19}: the relevant
exponent $(n+5/2)$ should be $7/2$ instead of unity.
\bibitem{25} M. Stobbe, Ann. Phys. (Leipzig) {\bf 7}, 661 (1930); H. Hall, Rev. Mod. Phys. {\bf 8}, 358 (1936).
\bibitem{26} H.A. Bethe and E. Salpeter, {\it Quantum Mechanics of One-and Two-Electron Atoms} (Plenum, New York,
1977).
\bibitem{27} L.D. Landau and E.M. Lifshitz, {\it Quantum Mechanics} (Pergamon, Oxford, 1965).
\bibitem{28} I.S. Gradshteyn and I.M. Ryzhik, {\it Table of Integrals, Series and Products} (Academic Press, New York,
1980).
\end{thebibliography}
\end{document}